\documentclass[useAMS,usenatbib,usegraphicx]{mn2e}
\usepackage{amsmath,epsfig,graphicx,color}
\title[X-ray and Radio from SN 2005kd]{X-Ray and Radio Emission from the Luminous Supernova 2005kd}
\author[Dwarkadas, Romero-Ca{\~n}izales, Reddy, \& Bauer]{V. V. Dwarkadas$^{1}\thanks{E-mail: vikram@oddjob.uchicago.edu~(VVD)}$, C. Romero-Ca{\~n}izales$^{2,3,5}$, R. Reddy$^{1}$,  and F. E. Bauer$^{3,2,4}$\\
$^{1}$Department of Astronomy and Astrophysics, U Chicago, 5640 S Ellis Ave, Chicago, IL 60637\\
$^{2}$Millennium Institute of Astrophysics,  Chile\\
$^{3}$ Instituto de Astrof\'{\i}sica, Facultad de F\'{\i}sica, Pontificia
Universidad Cat\'olica de Chile, Casilla 306, Santiago 22, Chile\\
$^{4}$Space Science Institute, 4750 Walnut Street, Suite 205, Boulder, Colorado 80301\\
$^{5}$N\'ucleo de Astronom\'{\i}a de la Facultad de Ingenier\'{\i}a, Universidad Diego Portales, Av. Ej\'ercito 441, Santiago, Chile}

\begin{document}
\newcommand{\vper}{\mbox{${v_{\perp}}$}}
\newcommand{\vpar}{\mbox{${v_{\parallel}}$}}
\newcommand{\uper}{\mbox{${u_{\perp}}$}}
\newcommand{\vperout}{\mbox{${{v_{\perp}}_{o}}$}}
\newcommand{\uperout}{\mbox{${{u_{\perp}}_{o}}$}}
\newcommand{\vperin}{\mbox{${{v_{\perp}}_{i}}$}}
\newcommand{\uperin}{\mbox{${{u_{\perp}}_{i}}$}}
\newcommand{\upar}{\mbox{${u_{\parallel}}$}}
\newcommand{\uparout}{\mbox{${{u_{\parallel}}_{o}}$}}
\newcommand{\vparout}{\mbox{${{v_{\parallel}}_{o}}$}}
\newcommand{\uparin}{\mbox{${{u_{\parallel}}_{i}}$}}
\newcommand{\vparin}{\mbox{${{v_{\parallel}}_{i}}$}}
\newcommand{\dout}{\mbox{${\rho}_{o}$}}
\newcommand{\din}{\mbox{${\rho}_{i}$}}
\newcommand{\da}{\mbox{${\rho}_{1}$}}
\newcommand{\mfast}{\mbox{$\dot{M}_{f}$}}
\newcommand{\mslow}{\mbox{$\dot{M}_{a}$}}
\newcommand{\beqn}{\begin{eqnarray}}
\newcommand{\eeqn}{\end{eqnarray}}
\newcommand{\be}{\begin{equation}}
\newcommand{\ee}{\end{equation}}
\newcommand{\noi}{\noindent}
\newcommand{\ftheta}{\mbox{$f(\theta)$}}
\newcommand{\gtheta}{\mbox{$g(\theta)$}}
\newcommand{\ltheta}{\mbox{$L(\theta)$}}
\newcommand{\stheta}{\mbox{$S(\theta)$}}
\newcommand{\utheta}{\mbox{$U(\theta)$}}
\newcommand{\xitheta}{\mbox{$\xi(\theta)$}}
\newcommand{\vs}{\mbox{${v_{s}}$}}
\newcommand{\ro}{\mbox{${R_{0}}$}}
\newcommand{\pa}{\mbox{${P_{1}}$}}
\newcommand{\va}{\mbox{${v_{a}}$}}
\newcommand{\vo}{\mbox{${v_{o}}$}}
\newcommand{\vp}{\mbox{${v_{p}}$}}
\newcommand{\vw}{\mbox{${v_{w}}$}}
\newcommand{\vf}{\mbox{${v_{f}}$}}
\newcommand{\lprime}{\mbox{${L^{\prime}}$}}
\newcommand{\uprime}{\mbox{${U^{\prime}}$}}
\newcommand{\sprime}{\mbox{${S^{\prime}}$}}
\newcommand{\xiprime}{\mbox{${{\xi}^{\prime}}$}}
\newcommand{\mdot}{\mbox{$\dot{M}$}}
\newcommand{\msun}{\mbox{$M_{\odot}$}}
\newcommand{\yr}{\mbox{${\rm yr}^{-1}$}}
\newcommand{\kms}{\mbox{${\rm km} \;{\rm s}^{-1}$}}
\newcommand{\lambdav}{\mbox{${\lambda}_{v}$}}
\newcommand{\lequ}{\mbox{${L_{eq}}$}}
\newcommand{\eqpratio}{\mbox{${R_{eq}/R_{p}}$}}
\newcommand{\ra}{\mbox{${r_{o}}$}}
\newcommand{\bfig}{\begin{figure}[h]}
\newcommand{\efig}{\end{figure}}
\newcommand{\tone}{\mbox{${t_{1}}$}}
\newcommand{\done}{\mbox{${{\rho}_{1}}$}}
\newcommand{\dsn}{\mbox{${\rho}_{SN}$}}
\newcommand{\dzero}{\mbox{${\rho}_{0}$}}
\newcommand{\ve}{\mbox{${v}_{e}$}}
\newcommand{\vej}{\mbox{${v}_{ej}$}}
\newcommand{\Mch}{\mbox{${M}_{ch}$}}
\newcommand{\mej}{\mbox{${M}_{e}$}}
\newcommand{\Mst}{\mbox{${M}_{ST}$}}
\newcommand{\dam}{\mbox{${\rho}_{am}$}}
\newcommand{\Rst}{\mbox{${R}_{ST}$}}
\newcommand{\Vst}{\mbox{${V}_{ST}$}}
\newcommand{\Tst}{\mbox{${T}_{ST}$}}
\newcommand{\no}{\mbox{${n}_{0}$}}
\newcommand{\Efif}{\mbox{${E}_{51}$}}
\newcommand{\rsh}{\mbox{${R}_{sh}$}}
\newcommand{\msh}{\mbox{${M}_{sh}$}}
\newcommand{\vsh}{\mbox{${V}_{sh}$}}
\newcommand{\vrev}{\mbox{${v}_{rev}$}}
\newcommand{\rpr}{\mbox{${R}^{\prime}$}}
\newcommand{\mpr}{\mbox{${M}^{\prime}$}}
\newcommand{\vpr}{\mbox{${V}^{\prime}$}}
\newcommand{\tpr}{\mbox{${t}^{\prime}$}}
\newcommand{\cone}{\mbox{${c}_{1}$}}
\newcommand{\ctwo}{\mbox{${c}_{2}$}}
\newcommand{\cthree}{\mbox{${c}_{3}$}}
\newcommand{\cfour}{\mbox{${c}_{4}$}}
\newcommand{\Te}{\mbox{${T}_{e}$}}
\newcommand{\Ti}{\mbox{${T}_{i}$}}
\newcommand{\Ha}{\mbox{${H}_{\alpha}$}}
\newcommand{\Rprime}{\mbox{${R}^{\prime}$}}
\newcommand{\Vprime}{\mbox{${V}^{\prime}$}}
\newcommand{\Tprime}{\mbox{${T}^{\prime}$}}
\newcommand{\Mprime}{\mbox{${M}^{\prime}$}}
\newcommand{\rprime}{\mbox{${r}^{\prime}$}}
\newcommand{\rfprime}{\mbox{${r}_f^{\prime}$}}
\newcommand{\vprime}{\mbox{${v}^{\prime}$}}
\newcommand{\tprime}{\mbox{${t}^{\prime}$}}
\newcommand{\mprime}{\mbox{${m}^{\prime}$}}
\newcommand{\Me}{\mbox{${M}_{e}$}}
\newcommand{\nh}{\mbox{${n}_{H}$}}
\newcommand{\rr}{\mbox{${R}_{2}$}}
\newcommand{\rf}{\mbox{${R}_{1}$}}
\newcommand{\vtwo}{\mbox{${V}_{2}$}}
\newcommand{\vout}{\mbox{${V}_{1}$}}
\newcommand{\dshell}{\mbox{${{\rho}_{sh}}$}}
\newcommand{\dwind}{\mbox{${{\rho}_{w}}$}}
\newcommand{\dslow}{\mbox{${{\rho}_{s}}$}}
\newcommand{\dfast}{\mbox{${{\rho}_{f}}$}}
\newcommand{\vfast}{\mbox{${v}_{f}$}}
\newcommand{\vslow}{\mbox{${v}_{s}$}}
\newcommand{\cc}{\mbox{${\rm cm}^{-3}$}}
\newcommand{\apj}{\mbox{ApJ}}
\newcommand{\apjl}{\mbox{ApJL}}
\newcommand{\apjs}{\mbox{ApJS}}
\newcommand{\aj}{\mbox{AJ}}
\newcommand{\araa}{\mbox{ARAA}}
\newcommand{\nat}{\mbox{Nature}}
\newcommand{\aap}{\mbox{AA}}
\newcommand{\gca}{\mbox{GeCoA}}
\newcommand{\pasp}{\mbox{PASP}}
\newcommand{\mnras}{\mbox{MNRAS}}
\newcommand{\apss}{\mbox{ApSS}}

\date{}

\pagerange{\pageref{firstpage}--\pageref{lastpage}} \pubyear{2002}

\maketitle

\label{firstpage}

\begin{abstract}
  { SN 2005kd is among the most luminous supernovae (SNe) to be
    discovered at X-ray wavelengths.  We have re-analysed all good
    angular resolution (better than $20''$ FWHM PSF) archival X-ray
    data for SN 2005kd.  The data reveal an X-ray light curve that
    decreases as t$^{-1.62 \pm 0.06}$.  Our modelling of the data
    suggests that the early evolution is dominated by emission from
    the forward shock in a high-density medium. Emission from the
    radiative reverse shock is absorbed by the cold dense shell formed
    behind the reverse shock.  Our results suggest a progenitor with a
    mass-loss rate towards the end of its evolution of $\ge$ 4.3
    $\times$ 10$^{-4} \msun \,{\rm yr}^{-1}$, for a wind velocity of
    10 km s$^{-1}$, at 4.0 $\times$ 10$^{16}$ cm. This mass-loss rate
    is too high for most known stars, except perhaps hypergiant stars.
    A higher wind velocity would lead to a correspondingly higher
    mass-loss rate. A Luminous Blue Variable star undergoing a giant
    eruption could potentially fulfill this requirement, but would
    need a high mass-loss rate lasting for several hundred years, and
    need to explain the plateau observed in the optical light
    curve. The latter could perhaps be due to the ejecta expanding in
    the dense circumstellar material at relatively small radii.  These
    observations are consistent with the fact that Type IIn SNe appear
    to expand into high density and high mass-loss rate environments,
    and also suggest rapid variability in the wind mass-loss
    parameters within at least the last 5000 years of stellar
    evolution prior to core-collapse.}
\end{abstract}


\section{Introduction}
\label{sec:intro}
The core-collapse of a massive star greater than about 8 $\msun$
results in a spectacular explosion, and the expansion of a very high
velocity shock wave into the ambient medium, leading to the formation
of a supernova (SN) of Type II or Type Ib/c. Type II SNe are further
divided into different subclasses, depending mainly on their optical
spectra or light curve.  Type IIn SNe form one of the more recent
subclasses of Type II SNe, having been first identified in 1990
\citep{Schlegel1990}. They are characterized by narrow lines on a
broad base in the optical spectrum \citep{Schlegel1990, kankareetal12,
  mauerhanetal13}. Surveys have revealed that they comprise between 1
and 4\% of the total core collapse SN population in a volume limited
sample \citep{eldridgeetal13}. Observations at optical
\citep{filippenko1997, taddiaetal13}, infra-red \citep{foxetal11},
X-ray \citep{ddb10,chandraetal12}, and radio \citep{chandraetal09b,
  chandraetal12} suggest that this class of SNe arises from
circumstellar interaction with a high-density
medium. Spectropolarimetric observations of luminous IIns
\citep{baueretal12} are consistent with circumstellar interaction, but
point to a complex origin for the various emission components.

There exists a wide diversity in SNe that exhibit IIn features, which
has greatly complicated the task of identifying their progenitors.
The prototypical IIns SN 1986J, SN 1988Z, and SN 1978K were only
observed years after explosion. More recently transient surveys have
found SNe that show IIn-like features very early in their evolution.
Despite over two decades of study, no consensus has been reached on
the identity of their progenitor stars, and it seems quite likely that
the class of IIns does not have one class of progenitors. Luminous
Blue Variable (LBV) stars have been suggested as the progenitors of
IIns \citep{galyam2009, smith10}, although this has been disputed by
some authors \citep{dwarkadas11b}.  Analysis of some IIns led
\citet{dwarkadas11b} to suggest that RSG stars that undergo high
mass-loss at the end of their lifetimes could be IIn progenitors,
while \citet{shr09} have suggested extreme RSGs such as VY CMa as Type
IIn progenitors. Thus, although many theories exist, identifying the
progenitors will require observations and analysis of a large sample
of Type IIns over the entire wavelength spectrum.

Supernova 2005kd was discovered in an automated search by \citet{pp05}
on 2005 Nov 12.22 UT.  It was confirmed by \citet{eastmanetal05} to be
a Type IIn SN. It lies in the galaxy LEDA 14370, at a redshift of
z=0.015040, which translates to a luminosity distance of about 63.2
Mpc (H$_0$=71 km s$^{-1}$ Mpc$^{-1}$, ${\Omega}_{\rm matter}$ = 0.27,
${\Omega}_{\rm vacuum}$ = 0.73).  X-ray and UV emission was observed
at the position of the SN with Swift in January of 2007
\citep{ipb07}. Following this, X-ray emission was detected with {\it
  Chandra} \citep{pif07}. The SN was also detected in the radio band
at 8.4 GHz \citep{cs07}. This is one of the furthest SNe to be
detected in X-rays, suggesting an unusually high X-ray luminosity. SN
2005kd was observed in the infra-red by the WISE satellite, however
the detection is questionable due to the proximity of the host nucleus
\citep{foxetal13,fox15}.

The optical lightcurve of SN 2005kd \citep{dyt08} shows an unusually
long plateau stage, unique for a Type IIn SN, which lasted for at
least 192 days. The UV lightcurves \citep{pritchardetal14} show an
even larger timespan where the flux appears to be either constant or
even increasing. Clearly, even for the strange class of Type IIn's
this represents a unique evolution.

In this paper we re-analyse and present all available X-ray data on SN
2005kd, combined with a 29 ks {\it Chandra} spectrum that we obtained
in Nov 2013. We use this data to study the evolution of the X-ray
light curve, the nature of the X-ray emission, the density structure
of the surrounding medium, and thereby the mass-loss properties of the
progenitor star. In \S \ref{sec:xraydata} we summarize the available
data, and describe the X-ray data reduction and fitting methods. \S
\ref{sec:xraylc} presents the X-ray lightcurve of SN 2005kd. In \S
\ref{sec:analysis} we interpret the lightcurve using semi-analytic
calculations, use it to extract information regarding the medium which
the SN is expanding in, and attempt to delineate the SN
progenitor. Motivated by the results from this section, in \S
\ref{sec:radio} we present available radio archival data that we have
reduced and analysed in order to improve our understanding of the SN
evolution. Finally, \S \ref{sec:disc} summarizes our work, and places
this SN in the context of other IIn SNe as well as the complete group
of core-collapse SNe.

\section[]{Data Analysis}
\label{sec:xraydata}
SN 2005kd was first detected in the X-ray band around 15 months after
explosion, and thus no data exist for the first year of evolution. We
have reduced all good angular resolution ($< 20'' $ FWHM PSF) data,
including data from {\it Swift}, {\it Chandra} and {\it XMM-Newton}.
The exposure times in general were short, such that much of the data
has low signal-to-noise. Table 1 summarizes the available data and
lists the derived fluxes.

The data from each satellite was reduced according to the standard
reduction procedures. X-ray spectral fitting was done using SHERPA
with thermal XSPEC models. We have presented results using the {\it
  vmekal} models, but note that using either {\it raymond} or {\it
  apec} models does not alter the derived conclusions. Given the low
number of data counts in all cases except for the {\it XMM-Newton}
pointing, all fits and flux estimations were done by fitting the
background first using a polynomial function, and then simultaneously
fitting the data plus the background, using the {\it cstat} statistic
on unbinned data. The high statistics of the {\it XMM-Newton}
observation allowed use of the {\it chi2gehrels} statistic on binned
data combined with background subtraction.  The flux estimates are
shown in Table 1.  Below we discuss the data reduction and analysis in
detail.

\begin{table*}
 \begin{minipage}{140mm}
  \caption{Summary of X-ray Data on SN 2005kd, listing the satellite
    which took the observation, the rest-frame corrected days after
    explosion, the exposure time, the column density and derived
    temperatures, and k-corrected unabsorbed fluxes, all with 1-$\sigma$ error
    bars where available. }
  \begin{tabular}{@{}llcrrrrrr@{}}
  \hline
   Satellite &  Obs Date  & Days After & Days After & Exposure  & Count Rate  & $N_H$ & kT & 0.3 - 8 keV \\
 & & Outburst\footnote{average value in case of combined exposures} & Outburst & (ks) & (10 $^{-3}$  & (10$^{22} {\rm cm^{-2}}$) & keV & Flux\footnote{using cstat statistic (except {\it XMM-Newton})} (10$^{-14}$ \\
& & & (Rest Frame) & & counts s$^{-1}$) & & & erg s$^{-1}$ cm$^{-2}$) \\
 \hline
 {\it Swift}    & 2007-01-24 & 440 & 433.5 & 8.9 & 3.9 $\pm$ 0.7 & 0.4 $\pm$ 0.27 & 17 $\pm$ 7 & 26$^{+13}_{-11}$\\
 {\it Chandra}  & 2007-03-04 & 479 & 472.0 & 3.0 & 22 $\pm$ 2.7 & 0.77 $\pm$ 0.17 & $>$ 20 & 49.6$^{+27}_{-16.8}$\\
 {\it XMM\footnote{average MOS1, MOS2}  }    & 2007-03-29 & 504 & 496.5 & 54.2 & 15 $\pm$ 0.7 & 0.94 $\pm$ 0.29 & $>$ 30 &  41.4$^{+4.1}_{-9.4}$\\
 {\it Chandra}  & 2008-01-03 & 784 & 772.4 & 5.0 & 17 $\pm$ 1.8 & 0.95 $\pm$ 0.49 & 6 $\pm$ 0.1 &  44.6$^{+25.}_{-25.3}$\\
 {\it Swift }   & 2008-08-21 & 1015 & 1000.0 & 9.3 & 2.2 $\pm$ 0.5 & 0.75\footnote{assumed} & 4.9$_{-2.3}^{+35}$ & 19.86$^{+1.87}_{-6.93}$\\
 {\it Swift}    & 2011-10-22 & 2200 & 2167.4 & 9.9 & 0.96  & 0.4$^{d}$ & 3.5$^{d}$  &$<$ 6.7 \\
          & to 2012-01-05 &  &     &     &   & & & \\
 {\it Swift}    & 2012-06-01 & 2419 & 2383.0 & 16.5 & 0.7  $\pm$ 0.3 & 0.15\footnote{converges to minimum set value (Galactic value)}$_{min}^{+0.15}$ & 4.4\footnote{upper bound unconstrained}$_{-2.9}^{\,[+75.5]}$ & 3.35$^{+0.18}_{-1.65}$\\
          & to 2012-07-16 &    &       &      &                      & & \\
 {\it Chandra}  & 2013-11-29 & 2940 & 2896.5 & 29.0 & 2.4 $\pm$ 0.4 & 0.43$_{-0.28}^{+0.12}$ & 3.12$^{+30}_{-0.4}$ & 1.98$^{+0.44}_{-0.36}$\\
 \hline
\label{table:05kddata}
\end{tabular}
\end{minipage}
\end{table*}

\subsection{XMM-Newton} 
\begin{figure}
\includegraphics[width=0.5\textwidth]{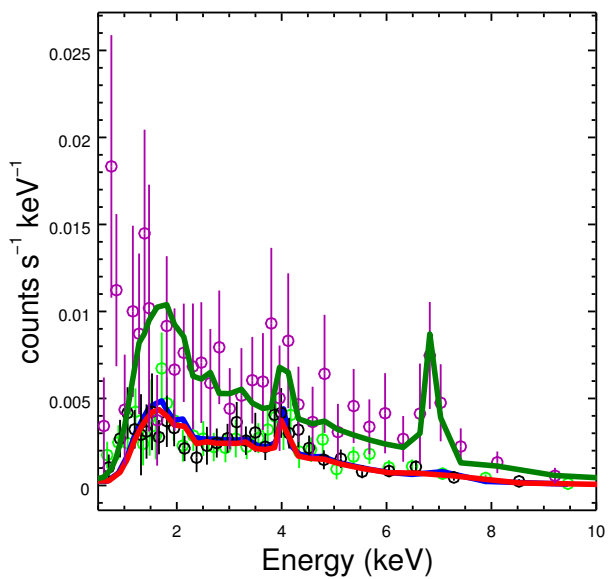}
\caption{{\it XMM-Newton} MOS1, MOS2 and PN data and fits. All
  datasets were fitted simultaneously using the same thermal {\it
    vmekal} model, and required abundances of Ca, Ar and Fe in excess
  of solar values. MOS1 data is in light green, and the corresponding
  fit in blue. MOS2 data is in black, and the fit in red. The PN data
  is in purple with the fit in dark green. The similarity in the fits
  demonstrates that the same model adequately fits all datasets
  simultaneously. The Fe-K$\alpha$ line at 6.7 keV is visible in the
  PN data alone, a result of the effective area differences between
  the MOS and PN detectors at 6.7 keV.
\label{fig:xmmall}}
\end{figure}

The SN was observed at an age of 500 days using {\it
  XMM-Newton}. This is the longest exposure, with by far the best
spectral resolution and effective area, and was used as a template for
the subsequent data reduction. The $> 600$ source counts allow for a
reliable fitting of the spectrum, providing the model template that we
adopted to fit all the other spectra. A 25\arcsec~region was used for
the source, with the background region being of the same size several
arcseconds away. The data were appropriately filtered and spectra
obtained using version 14 of the XMM-SAS software, following standard
data reduction procedures. The {\it Chandra} SHERPA software
\citep{sherpa01} was used for analysis and fitting of all spectra.
All the three datasets were fitted simultaneously to ensure the
tightest constraints.  Figure \ref{fig:xmmall} shows the spectra from
the MOS1, MOS2 and PN instruments on board XMM, together with a
thermal {\it vmekal} model that was used to jointly fit the three
spectra. A Ca line at 4 keV can be seen. The presence of lines
indicated that a thermal model was appropriate. In the fits we find
that allowing Ca, Ar, and Fe to deviate from solar values allows for
the best fit. These elements were allowed to float freely but linked
to have the same values in the MOS spectra, with only the
normalization varying freely in each case.  As can be seen from Figure
\ref{fig:xmmall}, the best-fit normalization is quite similar for MOS1
and MOS2. In general with all the fits, it is clear that allowing the
$\alpha$-element values to float freely tends to improve the fit. The
best-fit abundances in both the MOS and PN fits were found to be
higher than solar, with Fe and Ar having enhanced values $>$ 10 (in
terms of \citet{ag89} solar values), and Ca showing even larger
abundances but with an equally large 1-$\sigma$ variation: 280 $\pm$
279.  The MOS fits did not change appreciably with S thawed, but the
PN fit demanded values of S $>$ 10.  The best-fit temperature is high,
beyond the measurable range of {\it XMM-Newton}. In each case we have
used a single temperature model, which seems to match the spectra
well.  The best fit column density is found to be 9.4 $\pm 0.29 \times
10^{21}$ cm$^{-2}$, consistent with the trend seen earlier.
Unabsorbed fluxes are listed in Table \ref{table:05kddata}. The error
on the flux was computed using the {\it Sherpa} {$\it sample\_flux$}
routine, which computes the flux (typically) 1000 times, taking the
variations in the parameters into account, and then determines the
average flux and 1-$\sigma$ error from the computed fluxes.

We also attempted to fit a two temperature model to the {\it
  XMM-Newton} data, tying the abundance values of the two components
together. The two temperature model results in a very low second
temperature component as it tries to mainly fit the low temperature
region, giving temperatures around 0.2 keV. It does not change the fit
appreciably, but increases the unabsorbed flux considerably, because
the high column density results in an unabsorbed flux at low
temperatures that significantly exceeds the absorbed flux. We do not
consider these models viable, and they do not improve the fit in the
high-temperature region. It is possible that allowing the abundance
values of each component to vary completely independently of the other
may provide a better fit. This also significantly increases the
parameter space and fitting time however. Our main purpose in this
paper is to get a reasonable fit in order to estimate the flux and
luminosity from the SN, and calculate the light curves. We have found
that the 1-component fits are adequate for this purpose. The fluxes so
obtained are consistent within the error bars with those listed in the
{\it XMM-Newton} Serendipitous Source Catalogue, 3XMM-DR4, further
validating our choice of model and fits. Finally, some of the
assumptions inherent to our model, such as ionization equilibrium, are
validated in \S \ref{sec:analysis}.

\subsection{Swift} SN 2005kd was observed several times between 2007 and
2012 with the {\it Swift} X-ray Telescope (XRT), which has a
23\arcmin$\times$23\arcmin~field-of-view, 18\arcsec~FWHM PSF (although
the 90\% encircled energy diameter is $\approx$43\arcsec with mild
energy dependence), and provides spectra over the 0.3-8.0 keV band
with a spectral resolution of $\approx$0.1-0.2 keV.  Processed data
were retrieved from the {\it Swift} archive. A 35\arcsec region was
used to extract the data, with the background obtained from a larger
annulus region of 35-100\arcsec. Analysis was performed using FTOOLS
and custom software. We filtered for event grades 0-12 only, used the
standard XRT response matrix, and generated a position-dependent
ancillary response file using the FTOOL program {\it xrtmkarf}, which
provides nominal vignetting and PSF aperture corrections. Spectra
close enough in time were binned together to provide sufficient
signal-to-noise, as listed in Table \ref{table:05kddata}.

\begin{figure*}
\includegraphics[scale=0.95, angle=0]{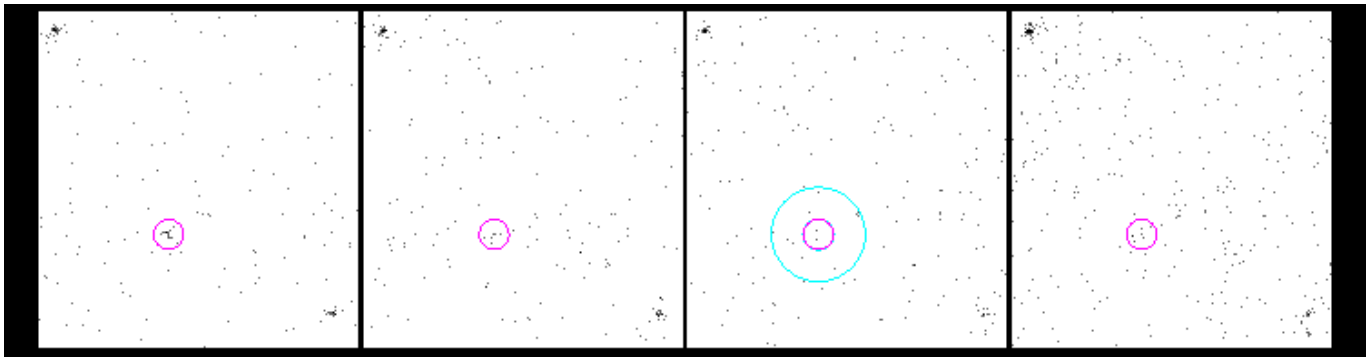}
\caption{Images from the {\it Swift} observations of SN 2005kd, in the
  0.3-8 keV energy range.  From left to right: 2007, 2008, 2011 and
  2012. The region is 11.5\arcmin~on each side.  The pink circle
  (35\arcsec~in radius) denotes the region used to extract the data
  for SN 2005kd. The annulus used for the background region is shown
  in blue in the 2011 panel. Clearly, the SN is not detected in 2011,
  but it is detected in all the other frames.
\label{fig:05kdswift} }
\end{figure*}

Figure \ref{fig:05kdswift} shows images from 4 different epochs during
which {\it Swift } data were obtained. The SN was detected in 2007 and
2008 with a significance $> 6 \sigma$. In 2012 it was detected again
but the significance was low ($\approx 3.3 \sigma$), prompting us to
request a follow-up {\it Chandra} observation.  The {\it XMM-Newton}
and {\it Chandra} spectra reveal that the spectrum is clearly thermal
in origin. Consequently for those epochs where the SN was detected, we
fitted the spectra using a single {\it vmekal} model.  We note that
the flux obtained for the 2007 dataset is almost identical to that
obtained by \citet{ipb07}. Our error bars are larger, which is
understandable given that we actually fit the spectrum, rather than
assuming a priori a thermal plasma with a temperature kT=10 keV, and a
Galactic N$_H$ of 1.5 $\times 10^{21}$ cm$^{-2}$ \citep{dl90}.  The
best-fit temperature is high (17 $\pm 7$ keV). The column density is
also higher than that used by \citet{ipb07}, but due to the higher
temperature tends not to have as much of an effect on the flux. We
find that the column density in the 2007 dataset is 4.0$\pm 2.8 \times
10^{21}$ (over twice the Galactic value).

The 2008 dataset initially gives a very high value of the column
density ($> 4.0 \times 10^{22}$ cm$^{-2}$), which results in an
extremely high value of the unabsorbed flux ($> 10^{-12}$ erg s$^{-1}$
cm$^{-2}$), using either statistic.  This value seems quite out of
place with the previously obtained values, suggesting a sudden
increase in column density, whereas the general trend seems to be
slowly decreasing. However, plotting the variation of the column
density versus temperature in the background fitted model shows that the
1-$\sigma$ range of possible values is large. By fixing the column to
a much smaller value of 7.5 $\times 10^{21}$ cm$^{-2}$ (intermediate
between the two nearest values, but closer to the previous {\it
  Chandra} value), we find an almost equally good fit, with a flux
that is an order of magnitude smaller. We have chosen to use this
lower flux, which is more appropriate. We can reduce the column
somewhat more, but the effect on the flux is within the error bars.

In 2011, the SN was not detected by {\it Swift} during an aggregate
9.3 ks XRT observation. Upper limits to the flux were calculated using
the Bayesian method of \citet{kbn91} for 99\% confidence. The uniform
prior used by these authors results in fairly conservative upper
limits, and other reasonable choices of priors do not materially
change our scientific results.  Within 35\arcsec, there are 4 counts,
of which 3 are expected to be background based on a larger annulus
region of 35-100\arcsec, once we scale by the source-to-background
area difference. Using the \citet{kbn91} upper limits for this source
+ background, we derive a 99\% upper limit of 8.9 counts. With an
exposure of 9.3 ks, this equates to 9.6 $\times 10^{-4}$ counts
s$^{-1}$. Using the Chandra PIMMS calculator, we deduce an upper limit
of 4.0 $\times 10^{-14}$ erg s$^{-1}$ cm$^{-2}$. This upper limit is
for the observed or absorbed flux. In order to get the unabsorbed
flux, we used the {\it Chandra} PIMMS calculator, assuming an
N$_H$=4.0$\times 10^{21}$ cm$^{-2}$ (again based on a decreasing N$_H$
within the 2008 and 2012 values) and a temperature of 3.5 keV
(intermediate between neighbouring values). A much lower temperature
could increase the flux by about 30-40\%, but we think this is a
better estimate given the parameters for the other observations
closest in time. This gives an unabsorbed flux limit of about 6.7
$\times 10^{-14}$ erg s$^{-1}$ cm$^{-2}$.

By 2012 the column density seems to have dropped to a value
approaching the Galactic value, the temperature has decreased, and the
flux is lower, as seen in Figure \ref{fig:05kdltcrv}. 

\subsection{Chandra} SN 2005kd was observed three times with the ACIS-S
instrument on {\it Chandra}. Observations of 3 and 5 ks were made in
2007 (ObsID 8518) and 2008 (ObsID 9095), respectively. Even in these
short observations, the SN was easily detected with $> 13 \sigma$
significance. After the SN was re-detected by {\it Swift} in 2012, we
proposed for, and were awarded a 30 ks observation, which was carried
out in Nov 2013 (ObsID 15999). The SN is detected with $> 7 \sigma$
significance, thus confirming the {\it Swift} re-detection.

The {\it Chandra} data were reduced and analyzed using the analysis
pipeline in the CIAO software version 4.7, and CALDB 4.6.8. A 4-arcsec
source region centered on the point source was used. Our spectra were
fitted with a thermal plasma {\it vmekal} model with a variable
temperature and column density. Given the results from the {\it
  XMM-Newton} fitting, we thawed elements Ca and Fe, as well as
Si. The C-statistic decreases, and the fit improves, when these
elements are thawed. We find abundance values of Fe and Ca similar to
those obtained with {\it XMM-Newton}, with Ca once again resulting in
highly elevated abundance values with large error bars.  For ObsID
8518, this gives a flux that is larger than the 2007 {\it Swift} value
by almost 80\%. We note that \citet{pif07} fitted a power-law model to
the data, found a column density comparable to the Galactic one and a
luminosity about 85\% higher than their {\it Swift} value, so these
results are quite consistent. Our best fit gives a column density 7.7$
\pm 1.7 \times 10^{21}$ cm$^{-2}$, which is higher than the Galactic
column density.  The best fit also suggests that the abundance of
metals such as Fe and Ca exceeds the solar value.

SN 2005kd was observed again by {\it Chandra} 10 months later. The
same thermal model was used as for the prior {\it Chandra} dataset. We
find the column density is 9.5$\pm 4.9 \times 10^{21}$ cm$^{-2}$, an
elevated Ca abundance exceeding solar, and temperature of $\sim$ 6
keV.

Finally, SN 2005kd was observed using ACIS-S on {\it Chandra} in Nov
2013. As mentioned before, the SN was detected with a high
significance, confirming its reappearance since 2011.  The best fit
thermal model gives a kT= 1.47$ \pm 0.39 $ keV, and N$_H$=4.3 $\pm
1.11 \times 10^{21}$ cm$^{-2}$, with a preference for an elevated Si
abundance. The column density is somewhat higher than the {\it Swift}
2012 value. The temperature is slowly decreasing as expected, since
the higher temperatures in the first 500 days.

\section{X-ray Lightcurve}
\label{sec:xraylc}

\begin{figure}
\includegraphics[width=0.5\textwidth]{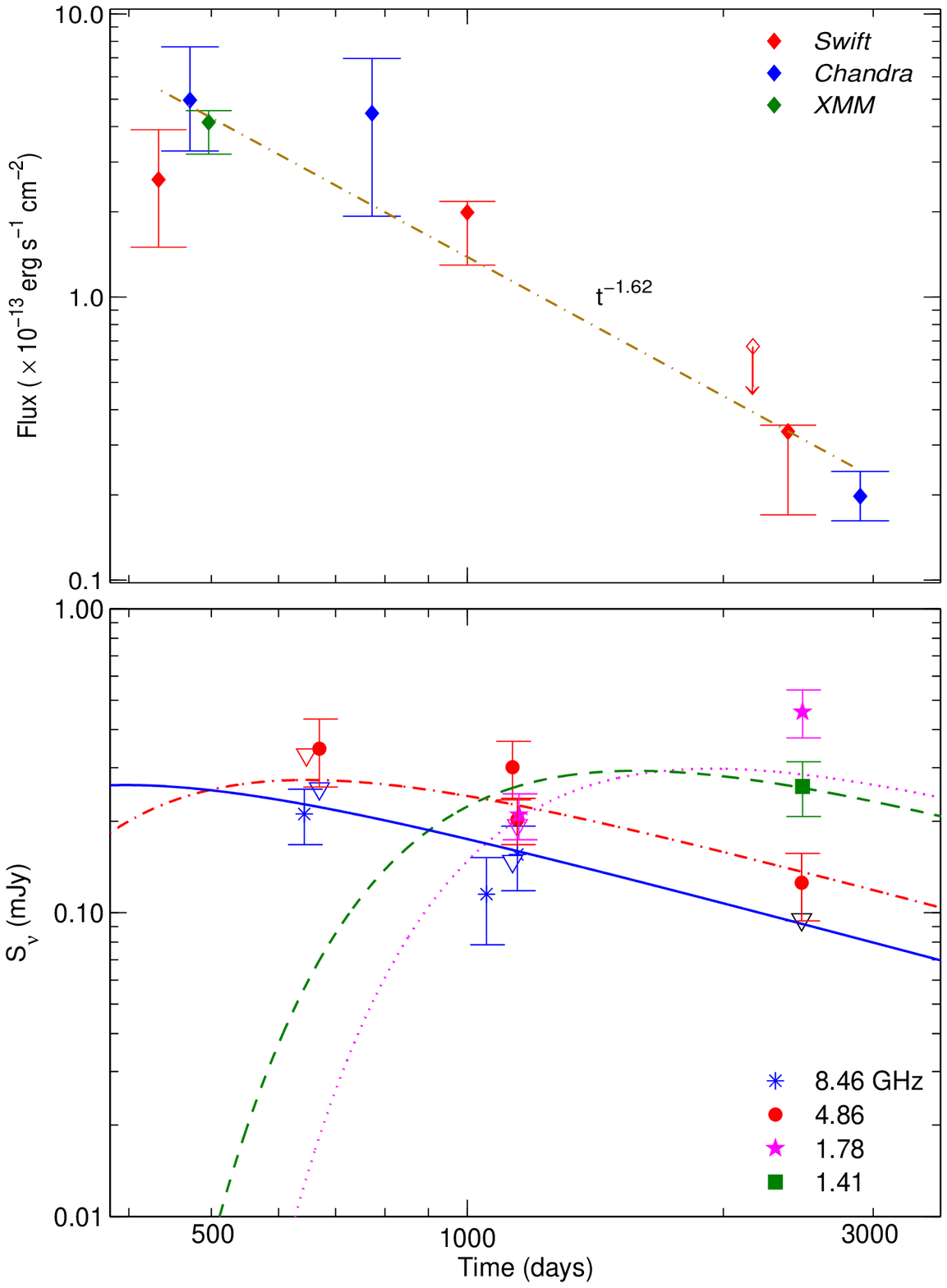}
\caption{({\bf Top}) The 0.3-8 keV X-ray lightcurve of SN 2005kd
  (epochs listed in Table 1), showing unabsorbed flux with 1$\sigma$
  error bars. The time is given in the rest frame of the SN, along
  with the k-corrected flux. The data comprise of 4 {\it Swift}, 3
  {\it Chandra} and 1 {\it XMM-Newton} data-points.  The best fit to
  the data is shown as a dashed line.  It suggests the flux is
  decreasing as $t^{-1.62 \pm 0.06}$.  ({\bf Bottom}) Radio lightcurve
  of SN 2005kd at 8.46 GHz (blue, solid line), 4.86 GHz (red,
  dashed-dotted line), 1.78 GHz (green, dashed line) and 1.41 GHz
  (magenta, dotted line).  Downward triangles represent 3$\sigma$
  upper limits at the respective frequencies according to their
  color. The black triangle is the upper limit at 7.91 GHz. The error
  bars represent $\pm 1\sigma$.  Where necessary, the flux has been
  converted to the appropriate frequency using the calculated radio
  spectral index. Fits to the data are shown and described further in
  \S \ref{sec:radio}. The few available data points do not allow for a
  robust fit, but the extracted parameter values are not inconsistent
  with those obtained from the X-ray data.}
\label{fig:05kdltcrv}
\end{figure}

Figure \ref{fig:05kdltcrv} (top panel) shows the complete X-ray
lightcurve of SN 2005kd calculated using background fitting. The data
are converted to the rest frame epoch, and k-corrected unabsorbed flux
is plotted. The fluxes are calculated as indicated above, with
1-$\sigma$ error bars. Subsequent to the 2008 {\it Swift} observation, for
over a 1000 days, there were no X-ray observations of SN 2005kd. We
have combined exposures from Oct 2011 to Jan 2012 to get a combined
almost 10 ks exposure with {\it Swift}. The SN is not detected at all
within this exposure. However it is in the next two exposures, with
{\it Swift} and {\it Chandra}.

\begin{figure*}
\includegraphics[scale=0.4, angle=90]{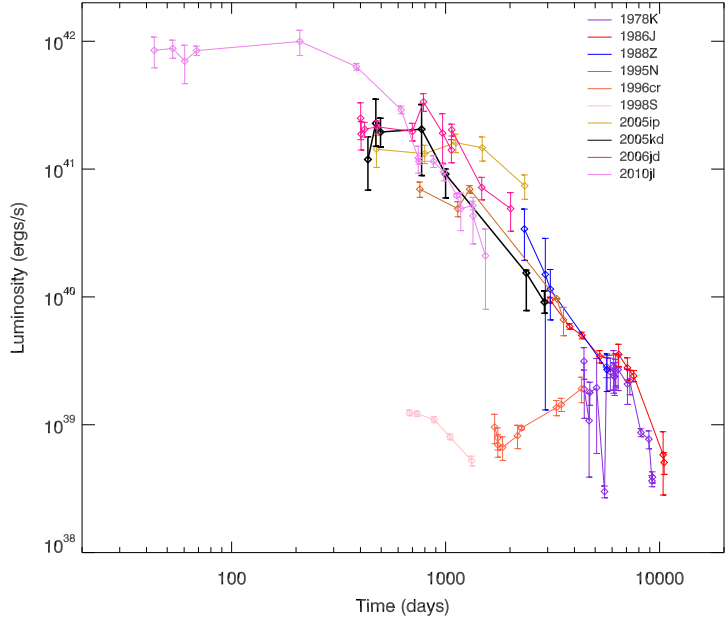}
\caption{The lightcurves of observed Type IIn X-ray SNe that have
  multiple exposures.  For information on the data for each SN, the
  X-ray fluxes, and the bands referenced for each SN, refer to
  \citet{dg12}, from which this figure has been adapted, with several
  additions.  SN 2005kd is shown with a thick black line. It's X-ray
  luminosity is high even among IIn SNe, which in general have the
  highest luminosities of all X-ray detected SNe. }
\label{fig:lcall}
\end{figure*}

SN 2005kd represents one of the most X-ray luminous Type IIns. Figure
\ref{fig:lcall} shows the lightcurve of other observed Type IIn
SNe. The X-ray lightcurve of SN 2005kd is shown as a thick black
line. The upper limit from the {\it Swift} 2011 observation is not
shown. It is clear that SN 2005kd is one of the most luminous, even
among Type IIn SNe, with a luminosity exceeding $ 10^{40}$ ergs
s$^{-1}$ over a period of about 550 days starting from day 440.  The
total energy deposition, in the 0.3-8 keV X-ray band alone, from days
400 to about 3000 is $> 10^{49}$ ergs. The total energy in X-rays is
likely to be much larger given the high temperatures inferred from
spectral modelling. This means that the SN radiated more than 1\% of
its kinetic energy within the first 3000 days at X-ray wavelengths,
unless the kinetic energy substantially exceeds 10$^{51}$
ergs. \citet{dyt08} had calculated a lower limit to the energy
radiated at optical wavelengths, finding it to be 3.2 $\times 10^{50}$
ergs in the first 500 days. Either the SN is radiating away a large
fraction of its energy in its first decade, or the total kinetic
energy exceeds the canonical 10$^{51}$ ergs by a substantial margin.

\section{Analysis and Interpretation}
\label{sec:analysis}
The overall lightcurve of SN 2005kd from 440 to 3000 days indicates a
decreasing luminosity with time. If we assume, as is generally the
case, that the light curve decays as a power-law in time, the best fit
to the data points gives a lightcurve that decreases as $t^{-1.62 \pm
  0.06}$ (Figure \ref{fig:05kdltcrv}). The temperature suggested by
the first 4 epochs is higher than the range of values that can be
measured by {\it Chandra}, {\it XMM-Newton} and {\it Swift}, and is
thus relatively unconstrained. The column density is higher than the
Galactic N$_H$ towards that direction (1.5 $\times 10^{21}$ cm$^{-2}$)
by a factor of 3-10, and appears to slowly decrease over time within
the (large) error bars. At $\sim$ 1000 days, the best fit spectral
model suggests a lower temperature and higher column density, but
plotting the 2D confidence contours of temperature vs.~column density
indicates that it is also compatible with the previous observations of
a high temperature and N$_H$ around 5 times the Galactic value.

The time evolution of the luminosity can be related to the density
structure of the surrounding medium, as shown in \citet{flc96} and
\citet{dg12}. To summarize, if we use the \citet{chevalier82a}
description for a SN shock wave evolving in a self-similar manner,
assuming spherical symmetry, the SN ejecta has a density that goes as
${\rho}_{SN} \propto v^{-n} \, t^{-3}$, and the uniform circumstellar
medium into which the SN evolves has a density profile that decreases
as ${\rho}_{CSM} \propto r^{-s}$, then the X-ray luminosity of the SN
will decrease as

\begin{equation}
L_x \propto t^{-(12-7s+2ns-3n)/(n-s)} \;\;\;\;\;
\label{eq:lcfull}
\end{equation}

or 

\begin{equation}
L_x \propto t^{-(6-5s+2ns-3n)/(n-s)} \;\;\;\;\;  E << kT_{sh}\;\; .
\label{eq:lcsm}
\end{equation}

\noindent
Equation \ref{eq:lcfull} is valid when one is considering the total
X-ray emission, equation \ref{eq:lcsm} when one is considering the
emission in an energy band $E$ where $E << kT_{sh}$, and $T_{sh}$ is
the shock temperature. Neither is an exact fit to this situation; the
latter is probably a better approximation over the first several
hundred days, the former over the later period, but the resulting
values are not significantly different using either equation. If we
consider a luminosity decreasing as $t^{-1.62 \pm 0.06}$, we find that
$s$ varies between 2.3-2.46 for $n = 9-12$.  Given the variation in
parameters, we assume $s=2.4 \pm 0.1$ in our analysis. This range of
$s$ indicates that the density decreases marginally faster than
$r^{-2}$, which would be the case for a wind with constant mass-loss
rate and wind velocity. We note from figure~\ref{fig:lcall} that this
steeper decrease is not uncharacteristic of Type IIn SNe at this age;
in fact many IIns appear to show a similar steep decrease in the
lightcurves.

In order to calculate the mass-loss rate, we follow the procedure
outlined in \citet{flc96}. Given the high temperature of the emission
over most of the first 1000 days, we assume that it arises from the
forward shocked circumstellar medium, and we do not see any emission
from the reverse shocked ejecta (this is addressed later).  We will
calculate the quantity $(\dot{M}/v_w)$ at a specific reference radius,
10$^{15}$ cm. We use the {\it XMM-Newton} observation given it has the
best statistics. Following \citet{flc96}, we use the spectral
luminosity at 1 keV. Since it is a spectral luminosity at a specific
temperature, it is not modified by line emission, unless the line
emission is present exactly at this frequency, which is not the case.
We write the luminosity of the forward (circumstellar) shock as
$L_{cs} \sim j_{ff}(T_{cs})M_{cs}{\rho}_{cs}/m_H^2$, where $j_{ff}\,
{\rho}_{cs}/m_H^2$ is the emissivity per unit mass, ${\rho}_{cs}$ is
the density behind the forward shock, $M_{cs}$ is the mass swept-up by
the forward shock, and $T_{cs}$ is the temperature behind the forward
shock.  The Gaunt factor $g_{ff}$ at 1 keV can be written as $g_{ff} =
1.87 \, T_8^{0.264}$, where $T_8$ is the forward shock temperature
$T_{cs}$ in terms of 10$^8$ K. This approximation does not deviate by
more than 30\% from more accurately tabulated values at each
temperature, as long as the energy $E < 15$ keV, but may not be as
appropriate for higher energies \citep{margon73}. Using this value of
$g_{ff}$, we can write the luminosity of the forward shock at 1 keV
as:

\begin{equation}
\begin{split}
L_{cs, 1keV} = 1.4 \times 10^{38} \,\xi \,T_8^{-0.236}
\frac{e^{-0.116/T_8}}{(3-s)}\left[\frac{{\dot{M}_{-5}}}{v_{w1}}\right]^2 \\
V_4^{3-2s}\,\left[\frac{t_d}{11.57}\right]^{3-2s} {\rm ergs \, s^{-1} \, keV^{-1}}\;\; .
\end{split}
\label{eq:lcs}
\end{equation}

\noindent
where $\dot{M}_{-5}$ is the mass-loss rate scaled to $10^{-5} \msun$
yr$^{-1}$, $v_{w1}$ is the wind velocity in terms of 10 km s$^{-1}$,
$V_4$ is the maximum ejecta velocity scaled to 10$^4$ km s$^{-1}$,
$\xi = [1 + 2n(He)/n(H)]/[1 + 4 n(He)/n(H)] \sim 0.85$ , and $t_d$ is
the time in days. We note that this expression has different
approximations from the previous ones, and is independent of $n$. For
$s=2.4$ it gives a flux decreasing as $t^{-1.8}$. It is close enough
given the other uncertainties.

At 500 days, the average {\it XMM} flux at 1 keV is
6.53$^{+2.13}_{-1.73} \times 10^{-14}$ ergs s$^{-1}$ cm$^{-2}$
keV$^{-1}$. Using a distance of 63.2 Mpc, and inserting in equation
\ref{eq:lcs}, with values of $0.6 < V_4 < 0.9$, $6 \le T_8 \le 9$, and
$2.3 < s < 2.5$ in equation \ref{eq:lcs} gives

\be 192 \le \left[\frac{\dot{M}_{-5}}{v_{w1}}\right] \le 656 \;.\ee

Therefore we deduce that, for a wind velocity of 10 km s$^{-1}$, the
mass-loss rate must be around (1.9-6.6) $\times 10^{-3} \msun$
yr$^{-1}$ at 10$^{15}$ cm.  If the wind velocity is higher, the
mass-loss rate is correspondingly higher. It is difficult to find a
progenitor that satisfies the velocity, mass-loss rate and light-curve
characteristics (discussed in further detail in \S \ref{sec:disc}). We
emphasize that the mass-loss rate is a time-varying quantity. However,
it is clear that the ambient medium around the SN has a high density.

As described in \S \ref{sec:xraydata}, we have assumed thermal models
to describe the SN, which are in ionization equilibrium. We can now go
back and confirm if that approximation is reasonable. We note that we
have calculated the mass-loss rate and wind velocity at 10$^{15}$ cm;
thus the electron density at 10$^{15}$ cm is:

\be
n_e = \left[\frac{1}{4 \pi \times\,10^{30}\,\xi m_H}\right]\left[\frac{\dot{M}}{v_{w}}\right] \sim (6.8-23) \times 10^9 \;\; {\rm cm}^{-3}
\label{eq:ne}
\ee

\noi where $m_H$ is the mass of a hydrogen atom.  At any given radius,
the density is $ (6.8-23) \times 10^9 (r_0/r)^s$ cm$^{-3}$, where $r_0
= 10^{15}$ cm. We assume that for ionization equilibrium, the
condition is that $n_e t \sim 10^{12}$ cm$^{-3}$ s \citep{sh10}.

At 500 days, assuming an average shock velocity of 6000-9000 km
s$^{-1}$ (appropriate for the high density), the shock radius is
2.6-3.9 $\times 10^{16}$ cm. We have $n_e t |_{500d} \sim (0.45-4)
\times 10^{14}$ cm$^{-3}$ s for $s=2.4$. At 1000 days, assuming a
slightly slower average velocity of 5000-8000 km s$^{-1}$, the shock
radius is 4.32 - 6.9 $\times 10^{16}$ cm, and the corresponding value
is $(2.26-23.6) \times 10^{13}$ cm$^{-3}$ s for $s=2.4$. For $s=2.5$
the density is somewhat lower, but still gives values $> 10^{12}$
cm$^{-3}$ s, whereas for $s=2.3$ the density is higher.  In all cases
we may conclude that the plasma is in ionization equilibrium at all
times during which the observations were taken. A lower velocity will
give a smaller radius and higher density so it will further reinforce
this argument. Thus we can justify our use of models for plasma in
ionization equilibrium.

Why do we not see (low-temperature) emission from the reverse shock,
given that the temperature behind this shock is lower, and more likely
to fall in the range of {\it Swift}, {\it XMM-Newton} and {\it
  Chandra}? Presumably it is because most of the emission from the
reverse shock is absorbed.  This is possible if the reverse shock is
radiative and a cool dense shell has formed behind it, which absorbs
the emission from the shock. The cooling time from the reverse shock
can be written as:

\be
\begin{split}
t_{cool, r} = 3.5 \times 10^9 \frac{(4-s)(3-s)^{4.34}}{(n-3)(n-4)(n-s)^{3.34}}\, \\ 
V_4^{3.34 + s} \left[\frac{\dot{M}_{-5}}{v_{w1}}\right]^{-1}\left[\frac{t_d}{11.57}\right]^{s} s \;\;.
\end{split}
\label{eq:rcool}
\ee

For $2.3 < s < 2.5$, $ 9 < n < 12$, and $V_4 < 1$ we find that the
reverse shock is radiative throughout the evolution, for the derived
values of $\left[\frac{\dot{M}_{-5}}{v_{w1}}\right]$.  Thus our
initial assumption of the emission arising from the forward shock is
validated. For large values of $n$, the cooling time goes as
$n^{-5.34}$ and thus is strongly dependent on the value of $n$. The
shock will be radiative for any larger value of $n$.

Within the error bars, we find that the column density is 3-10 $\times
10^{21}$ cm$^{-2}$, and decreases slowly over the first 1000
days. This suggests that there is some extra absorbing column density
ahead of the forward shock too. Assuming the ambient density extends
outwards continually with the same slope, the column density ahead of
the circumstellar shock can be written as

\be
N(H)_{cs} = 2.1 \times 10^{22} \left[\frac{\dot{M}_{-5}}{(s-1)\,v_{w1}}\right]\,V_4^{1-s}\left[\frac{t_d}{8.9}\right]^{1-s} {\rm cm}^{-2} \;\;.
\label{eq:column}
\ee

We note that this is the additional column that must be added to the
value of the Galactic column density. For parameters between
$\left[\frac{\dot{M}_{-5}}{v_{w1}}\right] \sim 192-656$, $0.6 < V_4 <
0.9$, $s=2.3-2.5$ and $t_d < 1000$, this value starts off as a few to
several times larger than the measured column of $\sim 3-10 \times
10^{21}$, and then slowly decreases, with the decrease being larger
for higher values of $s$ as expected from
equation~\ref{eq:column}. For values of $s$ = 2.4-2.5 and the lower
end of the mass-loss rate, the column density is generally less than
the Galactic column by about 3000 days, and does not contribute
much. For $s=2.3$, the value is still larger than Galactic even at day
3000.  For the top end of the mass-loss rate range, the values are
quite high, especially for $s$=2.3 and low $V_4$. The highest
mass-loss rates are less likely, since we do not see such a large
observed column, unless the medium is almost fully ionized (which is
not the case as we show below).  This implies that higher values of s,
combined with mass-loss rates at the lower end of their appropriate
range for high ’s’, are more probable.

The timescale for recombination is $\sim$ 3 $\times 10^{12}/ n_e$
s. Using $n_e$ from equation \ref{eq:ne}, we see that the
circumstellar material ionized by the progenitor star would have
already recombined by the time the shock wave reaches it \citep[see
also][]{vvd14}. Hence the main mechanism for ionization of the medium
is the X-ray emission itself, which depends on the ionization
parameter $\chi = L/nr^2$ \citep{km82}. This can be written as:

\be
\chi = 2 \times 10^{-38} \, L \,{\xi}^{-2}\left[\frac{\dot{M}_{-5}}{v_{w1}}\right]^{-1}\,V_4^{s-2}\left[\frac{t_d}{8.9}\right]^{s-2} \;\;.
\label{eq:ion}
\ee

Note that for $s=2$, this becomes independent of time (if $L$ is not
time dependent), as expected. For the luminosities of 2005kd, and the
high densities, the value of $\chi$ should be less than 100 throughout
most if not all of the evolution.  For high temperatures outside the
range probed by {\it Chandra} and {\it XMM-Newton}, e.g. $T_8 \sim
10$, \citet{ci12} find that elements such as C, N, and O are only
ionized when $\chi \sim 500$. This indicates that even the elements
like C, N, and O are not fully ionized in the high temperature
plasma. Heavier elements such as Fe are only ionized at $\chi \sim
5000$ \citep{ci12}. Overall the ionization factor is very low, which
suggests that the medium is mainly neutral, and we are seeing almost
the entire column that is present. The combination of
equations~\ref{eq:column} and ~\ref{eq:ion} indicates a preference for
higher values of $s$, which give higher ionization parameter and lower
column, that is more consistent with the data.

\section{Radio Emission from 2005kd} 
\label{sec:radio}

In an attempt to further constrain the evolution, we have investigated
the radio emission from the SN.  We have analysed SN\,2005kd radio
data at three different bands obtained with the Very Large Array
(VLA).  Data reduction was done following standard procedures using
the NRAO Astronomical Image Processing System ({\sc aips}) for data
taken in 2009 and earlier. For data taken after the VLA upgrade
completion, we used the Common Astronomy Software Applications package
\citep[{\sc casa};][]{mcmullinetal07}.

We used natural weighting in the imaging process of all the epochs to
increase their sensitivity.  Additionally there were observations at
22\,GHz made in November 2005 and August 2007, but no detections were
obtained. We do not discuss the emission at this frequency further
since the limits provide no additional constraints.

The observations included here were made using 0410$+$769 as a phase
calibrator. To test the reliability of the SN flux density variations,
we made maps of the phase calibrator and measured a peak intensity of
$\sim2.7$\,Jy/beam for all the 4.5\,GHz observations, and
$\sim1.8$\,Jy/beam at 8.5\,GHz. At 4.5\,GHz there is also a visible
source at about 1.2$^{\prime}$ West from SN\,2005kd.  This source has
a flux density $\sim1$\,mJy/beam in all those epochs. Thus, we
consider the flux densities shown in Table \ref{tbl:radio} as highly
robust measurements. The upper limits are given at a $3\sigma$
confidence level.

The SN was detected as a point source, hence, the peak intensities
shown in Table \ref{tbl:radio} also represent flux densities. The
uncertainties include the contribution of the local r.m.s. and a
conservative uncertainty in the absolute flux calibration of 5 per
cent. We note that the 2007 August 14 epoch corresponds to the radio
discovery reported by \citet{cs07}. Their value and the one we report
here are consistent within the uncertainties. Furthermore, the epochs
previous to the VLA upgrade in 2009 are limited by a poor dynamic
range owing to the existence of other sources in the field which are
much stronger than the SN itself. To account for this effect, we have
carefully measured any background emission at the position of the SN
in each epoch.

SN 2005kd was detected in the L, C and X bands over the first 9
years. The C and X band flux densities are decreasing with time,
whereas the L band light curve appears to be still rising. Thus the
radio emission appears to have already peaked at the higher
frequencies and transitioned from the optically thick to the optically
thin regime at 4.86 and 8.46\,GHz, whereas it is perhaps still in the
optically thick phase, or just transitioning to the optically thin
phase at 1.4 GHz. This is consistent with observations of other radio
SNe \citep{weiler02}. On 2012 August, the two-point spectral index
between 4.49 and 1.78 GHz is $\alpha = -0.72 \pm 0.35$. Using this
information, and assuming that the flux varied little between August
16 and 22, and that the spectral index extends from 1.78 to 4.86\,GHz,
we have converted the flux density measured on 2012 August 16 at 4.49
GHz to a flux at 4.86\,GHz.  The corresponding value is
$125.37\pm31.35\mu$Jy/beam.

We have fitted (see Figure \ref{fig:05kdltcrv}) a multi-frequency
light curve to the data shown in Table \ref{tbl:radio}, following the
parametrisation described in \citet{weiler02}, using a Monte-Carlo
simulation to obtain a robust fit \citep[see details
  in][]{canizalesetal14}.  We have adopted $\alpha = -0.72$ and
$t_0=$2005-Nov-10 as explosion date \citep{dyt08}. The optically thick
region is generally attributed to absorption by the external medium or
the medium internal to the SN (free-free absorption, FFA) or due to
synchrotron self-absorption (SSA). SN\,2005kd displayed radio emission
at late stages in its evolution, and a high but non-relativistic
ejecta velocity (based on the X-ray temperatures in the first 1000
days). Hence, we exclude the contribution of a clumpy CSM \citep[see
  e.g.][]{vandyketal94}, as well as synchrotron self-absorption
\citep{chandraetal12, canizalesetal14} and consider FFA in the uniform
local CSM. Thus, to represent the flux density evolution at a given
frequency, we have used:

\begin{equation}\label{eq:paramweiler}
\left(\frac{S}{1~\mathrm{mJy}} \right) = K_1\left(\frac{\nu}{5~\mathrm{GHz}} \right)^{\alpha}
\left(\frac{t-t_0}{1~\mathrm{day}}\right)^{\beta}e^{-\tau_{\mathrm{CSM}}} 
\end{equation}
where
\[ \tau_{\mathrm{CSM}}=K_2\left(\frac{\nu}{5~\mathrm{GHz}} \right)^{-2.1}\left(\frac{t-t_0}{1~\mathrm{day}}\right)^{\delta} \] \;\;.
The fitted parameters ($K_1=13.5$--$148.0$, $K_2=2.4\times
10^4$--$7.4\times 10^7$, $\beta=-0.74_{-0.17}^{+0.16}$ and
$\delta=-2.36_{-0.66}^{+0.46}$) allow us to infer a peak luminosity at
4.86\,GHz of 1.29$\times 10^{27}$ ergs s$^{-1}$ on day 631 after
explosion (rest frame). We thus estimate a mass loss rate of
$\sim0.5\times10^{-4}\,\msun$ yr$^{-1}$ \citep[assuming a wind
  velocity of 10\,km\,s$^{-1}$ and using equation 17 from][]{weiler02}
at a radius of $\sim4.6\times10^{16}$\,cm. The fit gives a value
$s=1.82\pm 0.37$. This slope is a bit lower than our X-ray derived
values. It is clear that there are problems with both the fit and its
interpretation. The former is due to the sparsity of data, which makes
fitting unreliable; the latter is mainly due to the complexity of the
system.

\begin{table*}
\centering
\begin{minipage}{180mm}
\caption{\protect{VLA observations of SN\,2005kd}}\label{tbl:radio}
\begin{tabular}{ccclcc} \hline
 \multicolumn{1}{c}{Observation}  & \multicolumn{1}{c}{Days after}  &\multicolumn{1}{c}{$\nu$} & \multicolumn{1}{c}{Conv. beam}
  &  \multicolumn{1}{c}{r.m.s.} &  \multicolumn{1}{c}{Peak intensity} \\
 \multicolumn{1}{c}{Date} & \multicolumn{1}{c}{outburst (rest frame)} & \multicolumn{1}{c}{(GHz)} & \multicolumn{1}{c}{(arcsec$^2$)}
  &  \multicolumn{1}{c}{($\mu$Jy/beam)} & \multicolumn{1}{c}{($\mu$Jy/beam)} \\
\hline
2007-Aug-14  &  632.5 & 8.46 & 0.38$\times$0.29, $ -42.7^{\circ}$ & 42.42 & 211.06$\pm$43.71 \\
2007-Sep-10  &  659.1 & 8.46 & 0.37$\times$0.29, $  40.3^{\circ}$ & 80.19 & $<$255.85        \\
2008-Sep-27  & 1036.4 & 8.46 & 0.37$\times$0.26, $ -35.9^{\circ}$ & 36.24 & 115.09$\pm$36.69 \\
2008-Dec-13  & 1112.3 & 8.46 & 0.39$\times$0.24, $ -36.5^{\circ}$ & 47.88 & $<$147.47        \\
2008-Dec-28  & 1127.1 & 8.46 & 0.36$\times$0.24, $ -20.4^{\circ}$ & 36.37 & 155.36$\pm$37.19 \\
2012-Aug-16  & 2434.4 & 7.91 & 1.47$\times$0.89, $  -2.2^{\circ}$ & 31.10 & $<$94.92         \\
\hline
2007-Aug-18  &  636.4 & 4.86 & 0.62$\times$0.44, $ -37.6^{\circ}$ & 82.41 & $<$331.13        \\
2007-Sep-10  &  659.1 & 4.86 & 0.58$\times$0.44, $  38.2^{\circ}$ & 85.56 & 346.57$\pm$87.30 \\
2008-Dec-13  & 1112.3 & 4.86 & 0.93$\times$0.38, $ -42.3^{\circ}$ & 64.04 & 300.91$\pm$65.78 \\
2008-Dec-28  & 1127.1 & 4.86 & 0.74$\times$0.36, $ -33.0^{\circ}$ & 33.56 & 202.54$\pm$35.05 \\
2012-Aug-16  & 2434.4 & 4.49 & 2.57$\times$1.57, $  -2.0^{\circ}$ & 32.53 & 132.80$\pm$33.20 \\
\hline
2008-Dec-28  & 1127.1 & 1.42 & 2.66$\times$1.41, $ -45.3^{\circ}$ & 47.71 & $<$193.57        \\
2009-Jan-02  & 1132.0 & 1.42 & 2.52$\times$1.37, $  -6.7^{\circ}$ & 34.54 & 209.94$\pm$36.10 \\
2012-Aug-22  & 2440.3 & 1.41 & 6.71$\times$5.18, $-178.2^{\circ}$ & 79.14 & 458.04$\pm$82.39 \\
2012-Aug-22  & 2440.3 & 1.78 & 5.32$\times$4.08, $   1.1^{\circ}$ & 51.70 & 269.40$\pm$53.31 \\
    \hline
   \end{tabular} 
 \end{minipage}
\end{table*}

The smooth decline of the radio emission at both 4.86 and 8.46 GHz
suggests a fairly smooth transition from 1000 to 3600 days. However,
the long-lasting radio emission does favour the presence of an overall
high density CSM, which is also inferred from the X-rays.

\section{Discussion and Conclusions}
\label{sec:disc}

In the previous section we form an overall picture of Type IIn SN
2005kd. In our model the SN expands in a medium with a density slope
$s \sim 2.3-2.5$, with a value of
$\left[\frac{\dot{M_{-5}}}{v_{w1}}\right] \sim 192-656$ at a radius of
$10^{15}$ cm, and decreasing with time. The X-ray emission is likely
dominated by the forward shock, with a high temperature and a column
density that reflects the high mass-loss rate. The reverse shock
remains radiative, with all emission from the reverse shock being
absorbed by a presumed cool dense shell behind it. The data suggest a
slight, although not conclusive, preference for higher values of
$s=2.4-2.5$, and lower mass-loss rates. Further observations are
needed to confirm or refute this.

This analysis of the X-ray emission from a Type IIn SN confirms that
the SN expands into a very high density medium with density decreasing
somewhat faster than $r^{-2}$. The density is higher than those
encountered around the majority of stars. This is consistent with
other Type IIns, although as we see from Figure \ref{fig:lcall}, SN
2005kd's X-ray luminosity appears to exceed that of most Type IIn SN.
The instantaneous density at any given radius, ${\mdot}_r$, can be
written as: ${\mdot}_r \propto {\mdot}{(r/r_0)}^{2-s}$. For $s=2.4 \,
{\mdot}_r \propto {(r/r_0)}^{-0.4}$. Thus the mass-loss rate at 4.0
$\times$ 10$^{16}$ cm would be about 77\% lower than that at 10$^{15}$
cm. Note that this indicates that the mass-loss rate was likely higher
than 10$^{-4} \msun$ yr$^{-1}$ at 4.0 $\times$ 10$^{16}$ cm.

The density profile is likely somewhat steeper than an $r^{-2}$
density profile, indicating a variation in the wind parameters in the
years leading up to the SN explosion. If we assume that the wind
velocity was of order 10 km s$^{-1}$, then this variation occurred
over at least the last 5000-7000 years of stellar evolution,
suggesting that the progenitor star increased its mass-loss rate
(and/or decreased the wind velocity) a few thousand years before
explosion

It is hard to determine which progenitor best fits this analysis. The
high mass-loss rate of 0.43-1.5 $\times$ 10$^{-3} \msun$ yr$^{-1}$, at
4.0 $\times$ 10$^{16}$ cm, for a 10 km s$^{-1}$ wind, is too high even
for a red supergiant (RSG) star at the extreme high-mass end of
mass-loss rates \citep{mj11}. \citet{humphreysetal97} have shown that
the yellow hypergiant IRC+10420 may have undergone high mass-loss
episodes where it lost mass at a rate of 10$^{-3} \msun$ yr$^{-1}$ for
about a 1000 years. This is lower than the time period of high
mass-loss inferred by us for a wind velocity of $\sim$10 km s$^{-1}$,
but not significantly so.  It is possible that a hypergiant star with
a somewhat higher mass-loss rate may just about meet the required
characteristics at the lower end of the deduced mass-loss rate.  The
star must lose several solar masses of material in this period.  The
derived high Ca abundances are consistent with the finding that
supergiant atmospheres are rich in Ca-Al silicates
\citep{specketal00}.  A RSG or hypergiant progenitor is also
compatible with the fact that the optical lightcurve from 2005kd
showed a plateau for about 192 days \citep{dyt08}, indicating that it
may arise from a progenitor with a large stellar radius. In this
context it is interesting to note that, from an analysis of their
R-band light, it has been found that the SNe IIn population
statistically bears more similarity to the SN IIP population than the
SN Ic population \citep{haberghametal14}. The latter would be expected
for really massive progenitors. The overall behaviour of SN 2005kd is
reminiscent of other IIns that show photometric evolution similar to
IIPs, as outlined in \S \ref{sec:intro}. \citet{mauerhanetal13} have
defined a subclass of SNe called Type IIn-P to describe these SNe that
showed both IIn and IIP characteristics. The energy radiated by all of
these in the plateau phase must arise from circumstellar medium
interaction, as demonstrated for SN 2005kd. As compared to the IIn-Ps
however, SN 2005kd shows a much longer plateau duration, and does not
appear to decline as rapidly as the others at the end of the plateau
phase. Unfortunately, the lack of optical spectroscopy on SN 2005kd
precludes more detailed comparison.

If the surrounding wind velocity were assumed to be much higher than
10 km s$^{-1}$, the mass-loss rate would need to be correspondingly
higher, thus making it difficult to ascribe a known progenitor to the
SN. LBVs have been suggested as IIn progenitors. Their wind velocities
are about an order of magnitude higher than those of RSGs, and
consequently the required mass-loss rates would jump up by an order of
magnitude.  LBVs undergoing a giant eruption \citep{smith14} would be
needed to satisfy the required mass-loss rates, which would exceed 1
$\times$ 10$^{-3} \msun$ yr$^{-1}$ for LBV wind velocities at 4.0
$\times$ 10$^{16}$ cm. The high mass-loss rates would also have to be
sustained for several hundred years, which has generally not been
observed in LBVs. Importantly, any progenitor without a large stellar
radius as in RSGs would furthermore require a different explanation
for the plateau region in the optical light curve. This could perhaps
be a consequence of the SN being surrounded by a massive dense shell
into which the shock expanded after breakout \citep{dha16}. The shell
would have to exist from a radius of around 10$^{16}$ cm, given the
current observations.

We have interpreted the obvious high density suggested by the
observations as a high mass-loss rate (in equation \ref{eq:lcs}). This
however is not the only interpretation. A high density shell can also
be due to sweeping up of external material by a lower density wind
\citep{dwarkadas11b}. A similar model was found adequate to explain
the X-ray emission from SN 1996cr \citep{ddb10}. The decreasing light
curve makes this appear less likely in the present case, however the
lack of X-ray observations in the first 400 days makes it difficult to
entirely rule out the possibility.

Figure \ref{fig:lcall} shows the similarity between the X-ray
evolution of SN 2006jd and that of SN 2005kd. The X-ray luminosities
are comparable. \citet{chandraetal12}, in their interpretation of the
X-ray lightcurve from SN 2006jd, have approximated it as a simple
power-law decline, similar to our assumption here. Their derived
densities were in excess of 10$^6$ cm$^{-3}$ at 1000 days, comparable
to the values found herein. The slope that they found was less steep
than that found here, but overall the similarities between these two
Type IIns are obvious.

In summary, SN2005kd provides further confirmation that IIn SNe evolve
in high mass-loss rate winds
\citep{Smith2007,ofeketal07,cwvetal11,chandraetal12,franssonetal14}.
They often appear to show rapid changes in wind parameters near the
end of the star's lifetime. Such changes have also been postulated for
many other Type IIn SNe \citep[][and references within]{ddb10}.  The
high mass-loss rates could just about accomodate a hypergiant star as
a progenitor. An alternative possibility is that the surrounding wind
velocity, and consequently mass-loss rate, are higher, and the
progenitor is not a RSG that turned hypergiant but an LBV star
undergoing a giant eruption, as has been often suggested for IIns. The
latter though would require extremely high mass-loss rates to be
sustained for at least several hundred years.  Any progenitor model
must also be able to account for the change in wind parameters in the
years prior to the explosion, as well as the optically observed
plateau region. Further continual and consistent monitoring of Type
IIn SNe at all wavelengths, but particularly in the X-ray and radio
bands, is suggested if we are to unambiguously determine their
progenitors.

\section*{Acknowledgments}
We acknowledge the anonymous referee for diligently reading the paper,
and providing extremely valuable comments that have considerably
improved the manuscript. The scientific results reported in this
article are based on observations made by the {\it Chandra, Swift} and
{\it XMM-Newton} X-ray observatories, and by the Very Large Array,
NRAO Socorro, which is a facility of the National Science Foundation
operated under cooperative agreement by Associated Universities,
Inc. This research has made use of data obtained from the Chandra Data
Archive and the HEASARC archive, and software provided by the Chandra
X-ray Center (CXC) in the application packages CIAO, ChIPS, and
Sherpa. We gratefully acknowledge support from grant GO4-15075X,
provided by NASA through the Chandra X-ray Observatory center,
operated by SAO under NASA contract NAS8-03060 (VVD, FEB); and from
the NASA Astrophysics Data Analayis program grant NNX14AR63G awarded
to University of Chicago (VVD). We also acknowledge support from
CONICYT through FONDECYT grant 3150238 (CR-C) and from project
IC120009 ``Millennium Institute of Astrophysics'' (MAS) funded by the
Iniciativa Cient\'{\i}fica Milenio del Ministerio Econom\'{\i}a,
Fomento y Turismo de Chile (CR-C, FEB), from CONICYT-Chile grants
Basal-CATA PFB-06/2007 (CR-C, FEB), FONDECYT 1141218 (FEB), PCCI
130074 (FEB), ALMA-CONICYT 31100004 (CR-C, FEB), and "EMBIGGEN" Anillo
ACT1101 (CR-C,FEB).

\bibliographystyle{mn2e} \bibliography{paper}

\bsp

\label{lastpage}

\end{document}